\documentclass[12pt,preprint]{aastex}

\newcommand{\kms}{km\thinspace s$^{-1}$}
\newcommand{\Kkms}{K\thinspace km\thinspace s$^{-1}$}

\newcommand{\Msun}{M$_\odot$}
 
\begin{document}

\title{High Resolution CO and H$_2$ Molecular Line Imaging of a
Cometary Globule in the Helix Nebula\altaffilmark{1} }


\author{P. J. Huggins\altaffilmark{2}, T. Forveille\altaffilmark{3,4}, 
R. Bachiller\altaffilmark{5}, P. Cox\altaffilmark{6},
N. Ageorges\altaffilmark{7}, J. R. Walsh
\altaffilmark{8} } 

\altaffiltext{1}{Based on observations
carried out with the IRAM Plateau de Bure Interferometer. IRAM is
supported by INSU/CNRS (France), MPG (Germany) and IGN (Spain).}
\altaffiltext{2}{Physics Department, New York University, 4 Washington
Place, New York, NY 10003, USA}
\altaffiltext{3}{CFHT, PO Box 1597, Kamuela, HI 96743, USA}
\altaffiltext{4}{Observatoire de Grenoble, B.P. 53X, 38041 Grenoble
Cedex, France}
\altaffiltext{5}{IGN Observatorio Astron\'omico Nacional, Apartado
1143, E-28800 Alcal\'a de Henares, Spain}
\altaffiltext{6}{Institut d'Astrophysique Spatiale, Universit\'e de
Paris XI, 91405 Orsay, France}
\altaffiltext{7}{European Southern Observatory, Alonso de C\'{o}rdova
3107, Santiago, Chile}
\altaffiltext{8}{Space Telescope Coordinating Facility, ESO, D-85748
Garching bei M\"{u}nchen, Germany}



\begin{abstract}
We report high resolution imaging of a prominent cometary globule in
the Helix nebula in the CO $J=1-0$ (2.6~mm) and H$_2$ $v=1-0$ $S(1)$
(2.12~$\micron$) lines. The observations confirm that globules consist
of dense condensations of molecular gas embedded in the ionized
nebula. The head of the globule is seen as a peak in the CO emission
with an extremely narrow line width (0.5~\kms) and is outlined by a
limb-brightened surface of H$_2$ emission facing the central star and
lying within the photo-ionized halo.  The emission from both molecular
species extends into the tail region. The presence of this extended
molecular emission provides new constraints on the structure of the
tails, and on the origin and evolution of the globules.

\end{abstract}


\section{Introduction}

The cometary globules in the Helix nebula (NGC~7293) are among the
most remarkable structures seen in planetary nebulae (PNe). They occur
in large numbers in the lower ionization regions of the nebula, and
appear in high resolution optical images as small ($\sim 1\arcsec$),
convex, photo-ionized surfaces facing the central star, with
comet-like tails extending in the opposite direction (e.g., Meaburn et
al. 1992; O'Dell \& Handron 1996).  These structures are seen in other
PNe (e.g., O'Dell et al. 2002) and are probably quite common, but they
are best seen in the Helix nebula because it is the nearest example,
at a distance of $D\sim 200$~pc ($\rm{parallax} = 4.70 \pm 0.75$~mas,
Harris et al. 1997).

A key step in understanding the nature of the Helix globules has been
their detection in CO by Huggins et al. (1992).  With a resolution
$\sim 12\arcsec$, the CO observations did not resolve their structure,
but demonstrated that globules contain a major mass component of
molecular gas, consistent with observations of dust, seen in
absorption against the nebula emission by Meaburn et al. (1992).  The
molecular gas places important constraints on the origin and evolution
of the globules, and provides a direct connection with the massive
shell of neutral gas that surrounds the ionized nebula (Forveille \&
Huggins 1991; Young et al. 1999; Rodriguez, Goss, \& Williams
2002). This connection is underscored by wide field imaging of the
nebula in the infrared lines of H$_2$ by Kastner et al. (1996), Cox et
al. (1998), and Speck et al. (2002), at resolutions from 8\arcsec\ to
2\arcsec, that show a highly fragmented envelope.

In order to determine the detailed relation between the molecular gas
and the structure of the globules revealed by optical images, we have
made new observations of the Helix nebula in both CO and H$_2$ with
significantly better resolution than previous observations. In this
\emph{Letter}, we report results on the molecular gas in a single
cometary globule that resolve its head-tail structure.

%

\section{Observations}

The globule observed is a prominent feature lying within the ionized
nebula to the north of the central star at offsets ($-$10\arcsec,
+135\arcsec). It is designated C1 by Huggins et al. (1992) and its
location is shown in Fig.~3a of Meaburn et al. (1998) where it is
labeled ``1''.  Optical images of the globule from Walsh \& Meaburn
(1993) in H$\alpha$+[N\,{\sc ii}]\,6584\,\AA\ and in [O\,{\sc
iii}]\,5007\AA, where it is seen in absorption against the nebular
emission, are shown in the top right panels of Fig.~1.
 
The CO observations were made in the 2.6~mm (115~GHz) $J=1-0$ line
using the IRAM interferometer at Plateau de Bure, France, in May 1999.
The array consisted of five, 15~m antennas, equipped with SIS
heterodyne receivers. The observations were made with the D
configuration of the array, with maximum baselines of $\sim
147$~m. The primary beam size of the interferometer is 44\arcsec\ at
2.6~mm, and the effective velocity resolution for the analysis is
0.2~\kms.  The uv data were Fourier transformed and CLEANed, using the
Clark algorithm, and the restored Gaussian clean beam is $7\farcs9
\times 3\farcs8$, at a position angle of 14\degr. The results are
shown in Figs.~1 and 2. 

Images of the H$_2$ $v=1-0$ $S(1)$ emission in the northern quadrant
of the Helix nebula were obtained with the SOFI infrared camera on the
ESO NTT, in June, 2001. The instrument has a 1024$^{2}$ Hawaii HgCdTe
array and was used with an image scale of 0\farcs24 pixel$^{-1}$. The
seeing was {1}\farcs{2}. The observations were made with an H$_2$
filter of width 0.028~$\micron$, centered at 2.124~$\micron$, well
separated from the He\,{\sc i} 2$^{1}$P$_{0}$--2$^{1}$S 2.058~$\micron$  and
H\,{\sc i} 2.166~$\micron$ lines, the latter having a transmission of
a few percent in the filter passband. Fifteen 1~min exposures were
made, using sky offset positions at 5\arcmin~N, 5\arcmin~E, and
7\arcmin~NE. The individual images were sky cleaned, flat fielded, and
combined with shift-and-add registration using the brightest stellar
images.

Many globules are detected in the full H$_2$ image, and the region
around globule C1 is shown in the top, center-left panel in
Fig.~1. An approximate calibration has been made by comparing the
smoothed, full image with the data of Speck et al. (2002). The
intensity of the emission at the head of the globule is 
$\sim 10^{-4}$~erg\,s$^{-1}$\,cm$^{-2}$\,sr$^{-1}$.  Astrometry of the
H$_2$ and optical images was carried out using stars in the USNO
catalog, and the registration between the images is $\sim
{0}\farcs{1}$ rms. The accuracy of the absolute positions for
comparison with the CO interferometry is $\sim$
{0}\farcs{3}--{0}\farcs{5}.

\section{Properties of the Globule}
 
\subsection{Overview}
The observations presented in Fig.~1 provide complementary views of 
the Helix globule. The CO 1--0 line, with an upper level of $E_{\rm
u}=5.5$~K above the ground state, shows the overall distribution and
kinematics of the cool molecular gas; the high lying ($E_{\rm u}\sim
7,000$~K) H$_2$ $v=1-0$ $S(1)$ line traces excited molecular gas; the
image in [O\,{\sc iii}] shows the distribution of dust, seen in
absorption against the nebular emission; and the image in
H$\alpha$+[N\,{\sc ii}] shows the photo-ionized surfaces facing the
central star. 

The observations immediately confirm the molecular nature of the globule,
and reveal some important details of its structure.  

%

\subsection{CO Structure and Kinematics}
The CO 1--0 velocity-integrated intensity map (Fig.~1) is seen to be
extended with respect to the telescope beam, and the deconvolved
source size, assuming a Gaussian distribution, is $\sim 2\arcsec
\times 10\arcsec$ (1\arcsec\ corresponds to $3\times10^{15}$~cm at
200~pc).  The emission is marginally resolved in RA but is resolved in
Dec, and extends roughly along the head-tail axis of the globule. The
peak intensity is 2.6~K\,\kms, and the total flux is
166~K\,\kms\,arcsec$^{2}$.
 
The CO velocity-strip (Fig.~2) shows that to the south, toward the
head of the globule, the line width is extremely narrow, $\Delta V =
0.5$~\kms\ (FWHM), and it broadens out to $\sim 0.8$~\kms\ farther
north into the tail region.  The CO radial velocity is precisely
determined to be $V_{\rm lsr} = -27.9$~\kms\ (the correction to
$V_{\rm hel}$ is $-$2.9~\kms), consistent with $-28.7\pm2$~\kms\
measured in [N\,{\sc ii}] 6584\AA\ by Meaburn et al. (1998). The
systemic velocity of the whole envelope of the Helix nebula is
$-$23~\kms\ (Young et al. 1999), so the globule is blue-shifted by
$\sim 5$~\kms.  For a radial expansion velocity of the globule system
of $\sim 20$~\kms\ (Young et al. 1999), the globule lies on the near
side of the nebula, on a radius vector from the star inclined to the
line of sight by $\sim 75\degr$. The head-tail axis, assumed radial,
is also seen at this angle.

The CO channel maps (Fig.~1) show the structure in the molecular gas.
There are two peaks, offset 3\arcsec~N and 8\arcsec\,N from the field
center. The first is just north of the maximum absorption
in the [O\,{\sc iii}] image (centered at $1\farcs5$\,N), and we
identify this molecular emission with the head of the globule whose
photo-ionized surface toward the central star is seen in the
H$\alpha$+[N\,{\sc ii}] image. Substantial CO emission, however,
extends away from the head into the tail region, and the second CO
peak lies close to a second, weak maximum in the dust absorption image
(at 8\arcsec\,N). This extended CO emission, together with the line
broadening into the tail accounts for the offset of the peak in the CO
integrated intensity map from the head of the globule.

These observations demonstrate that the molecular gas in the cometary
globule is \emph{not} a compact spheroidal bullet, but 
has a substantial component extending into the tail region.
 
%

\subsection{H$_2$ Distribution and Excitation}

The distribution of the H$_2$ emission in the globule is strikingly
different from that of CO (see Fig.~1). The strongest H$_2$ emission
occurs at the face of the globule toward the central star. There is
little emission directly behind the globule, but observable emission
extends to large distances ($\gtrsim 24\arcsec$) along the
tail. Remarkably, the H$_2$ distribution most closely follows that of
the ionized gas seen in H$\alpha$+[N\,{\sc ii}].

The high spatial resolution of our observations and the well defined
geometry of the globule provide a unique perspective on the question
of the excitation of the H$_2$ (Cox et al. 1998), which affects its
observed distribution. The bright H$_2$ emission clearly arises in a
thin surface layer in the molecular gas.  Fig.~3 shows a close-up of
the globule head, comparing the H$_2$ and H$\alpha$+[N\,{\sc ii}]
emission with [O\,{\sc iii}].  The H$_2$ emission forms a
limb-brightened cap on the globule facing the central star, and lies
just inside ({0}\farcs{5}--1\arcsec) the halo of photo-ionized gas
seen in H$\alpha$+[N\,{\sc ii}].  In the tail region too (Fig.~1), the
H$_2$ is enhanced near two H$\alpha$+[N\,{\sc ii}] peaks, at
8\arcsec\,N and 12\arcsec\,N on the east side, which are directly
illuminated by the star. More detailed observations are needed to
determine whether these are separate, small globules along the line of
sight, or sub-structures of the main tail.

The observed distribution of the H$_2$ emission is in complete accord
with the expectations of H$_2$ excitation in a thin photo-dissociation
region (PDR) at the surface of the molecular gas. The observed
intensity of the line is probably consistent with this interpretation
(Natta \& Hollenbach 1998; Speck et al. 2002) although detailed PDR
models of the globules have not yet been developed. The possibility
that the observed distribution is caused by shocks is unlikely, in
view of the small crossing time ($\lesssim 400$~yr) for shocks
($v_{\rm s} \gtrsim 5$~\kms) able to excite the $v = 1-0$ $S(1)$ H$_2$
line (e.g., Burton et al. 1992); shocks in the bulk of the molecular
gas are ruled out by the absence of any disturbance in the CO emission
with velocities larger than $\sim 0.5$~\kms.

\subsection{Physical Properties of the Molecular Gas}

Using the CO 2--1 observations of the globule by Huggins et
al. (1992), we find the 2--1/1--0 flux ratio to be $\sim$ 2--3
(assuming significant 1--0 flux is not resolved out on the shortest
baselines), which suggests that the lines are at least partly
optically thin, with an excitation temperature $\sim$ 18--40~K in
the thin limit. For a representative value of 25~K and a CO abundance
of $3\times10^{-4}$, the CO 1--0 flux gives a mass of molecular gas in
the globule of $M_{\rm m} \gtrsim 1 \times 10^{-5}$~\Msun. The
corresponding average density in a volume with projected dimensions of
$2\arcsec \times 10\arcsec$, which includes the head and the 
tail seen in CO, is $n_{\rm H}\gtrsim 2\times 10^4$~cm$^{-3}$.

These values are consistent with a mass of $\sim 2 \times
10^{-5}$~\Msun\ and $n_{\rm H} \sim 4\times 10^5$~cm$^{-3}$ determined
for the head of the globule by Meaburn et al. (1992) from the dust
absorption seen in [O\,{\sc iii}], corrected to a distance of 200~pc.
The typical mass of photo-ionized gas at the surface of a globule,
$\sim 10^{-9}$~\Msun\ (e.g., O'Dell \& Handron 1996), is negligible in
comparison.

\section{Origins and Evolution}

The structure and kinematics of the molecular gas in the head and in the
tail of the globule provide basic constraints on their origin and
evolution.

One scenario for the origin of globules is that they form in the
atmosphere of the progenitor star and are carried out in the expanding
circumstellar envelope (Dyson et al. 1989). In this case, the
wind-swept appearance of the globules suggests a model in which
material from the head is swept into the tail by a radially directed
wind, and Dyson, Hartquist, \& Biro (1993) have shown that the wind
needs to be subsonic to form a narrow tail. Our observations constrain
this model in showing no evidence for a wind-swept flow pattern at the
present time. There is no difference in velocity between the CO
emission in the head and the tail of the globule, and from the CO
strip map (Fig.~2) the differential motion is $\lesssim 0.2$~\kms\
along the line of sight, or $\lesssim 0.8$~\kms\ in a radial
direction, corrected for the inclination. This would produce a tail of
length $\lesssim 4\arcsec$ in 5,000~yr which is likely the maximum
time available for such a process.  The tails could have been fully
formed at an earlier epoch, or it might be that ablation occurs
only from the ionized edges of the head of the globule, but this would
not account for the presence of molecular gas seen in H$_2$ along the
whole length of the tail.

A different view of the gas dynamics around the globule at the present
time is provided by the photo-evaporation model, which has been
discussed in the context of the Helix globules by
L\'{o}pez-Mart\'{\i}n et al. (2001). In this model, photo-ionization
of the neutral globule, whose molecular core is unambiguously
confirmed by our observations, produces an ionized out-flow
from the surface. Given that the density of the ionized gas at the
head of the globule is $\sim 10^{3}$~cm$^{-3}$, and that of the
ambient gas is much lower $\sim 50$~cm$^{-3}$ (e.g., O'Dell \& Handron
1996), any subsonic flow of the ambient gas around the globule is
unlikely to have important dynamical effects in shaping the gas in the
tail.

An alternative mechanism for growing a tail on a pre-existing globule
is by shadowing (Cant\'{o} et al. 1998), which can form a tail behind
the globule as the result of under-pressure of the gas in the shadowed
region. For the simplest case of a neutral globule in ionized gas,
this model produces only modest density increases in the tail, and is
unlikely to lead to the substantial molecular tails revealed by the
present observations.  Striking effects of shadowing of the ionizing
radiation are, however, seen in optical images of the Helix (e.g.,
Henry, Kwitter, \& Dufour 1999), where long, radial plumes appear as
extensions to globules and their tails. Similar shadowing of the
stellar radiation at wavelengths longer than the Lyman limit must also
occur, and the consequent reduction of the CO and H$_2$
photo-dissociation rates in the shadowed regions must play a key role
in preserving the molecular gas in the extended tails.

If globules originate close to the star as proposed by Dyson et
al. (1989), the most crucial phase in their evolution occurs when they
are overrun by the ionization front of the nebula. At this stage, the
shadowing of pre-existing molecules or hydrodynamic flows, or both,
could lead to globules with molecular tails. An alternative scenario
is one in which the globules and their tails originate simultaneously
in instabilities at the ionization front (e.g., Capriotti 1973). In
this model, shadowing and possibly hydrodynamic flows would also
likely play a role, and the fingers of gas which result from the
instability could contribute directly to the formation of the tails.
The later development of these two models will likely be similar, and
can plausibly lead to the globules with substantial molecular tails
that we now observe. Further studies should allow us to discriminate
between them.

As for the fate of the globules, Meaburn et al. (1998) estimated an
ablation rate for globule C1 of $3\times10^{-8}$~\Msun\,yr$^{-1}$
based on densities inferred from the [O\,{\sc iii}] image, and an
assumed ablation flow velocity of 10~\kms.  This implies an improbably
short lifetime of $\sim 500$~yr which led Meaburn et al. to consider
ad hoc, restricted flow models around the globule. The absence of an
observable ablation flow in the globule reported here for CO ($\lesssim
0.8$~\kms) completely resolves this problem, by increasing the original
estimate of the ablation time scale by at least an order of
magnitude. In fact, the current mass-loss rate of the globule is
likely dominated by photo-evaporation (L\'{o}pez-Mart\'{\i}n et
al. 2001), and the corresponding time scale is $\sim10^4$~yr. Thus the
globules are long lived, and may even escape the nebula.


\acknowledgements
We thank the staff of the Plateau de Bure interferometer for making
the observations.  This
work has been supported in part by NSF grant AST-9986159 (to P.J.H.).


\clearpage


\clearpage




\figcaption[]{Observations of the Helix globule C1. Top row: Images in CO
1--0 integrated intensity (left); H$_2$ $v=1-0$ $S(1)$ (center left);
H$\alpha$+[N\,{\sc ii}]\,6584\,\AA\ (center right); and dust
absorption, seen against the nebula emission in [O\,{\sc
iii}]\,5007\AA\ (right).  Bottom row: CO 1--0 channel maps; the
channels are 0.2~\kms\ wide and are centered at the velocities given
in the upper right of each panel.  The contour intervals are
0.4~\Kkms\ for the CO integrated intensity map, and 0.9~K for the
channel maps. The ellipse in the top, left panel shows the beam size
for the CO observations.  For each panel, the offsets are relative to a
field center of 22:29:37.78 $-$20:48:02.00 (J2000) which is used for
all images in the paper. }
\label{}

\vspace{0.5cm}

\figcaption[]{CO 1--0 velocity strip map of the globule along the major
axis.  The contour intervals are 0.5~K.
}
\label{}

\vspace{0.5cm}

\figcaption[]{Close-up of the head of the globule. Contours of the
emission in H$_2$ $v=1-0$ S(1) (left) and H$\alpha$+[N\,{\sc ii}]
(right), superposed on the [O\,{\sc iii}] image. The H$_2$ data are
smoothed with a Gaussian {0\farcs6} (FWHM). The contours are 0.3, 0.5,
0.7, and 0.9 of the peak emission.
}
\label{}

\newpage 
\begin{figure*}
\plotone{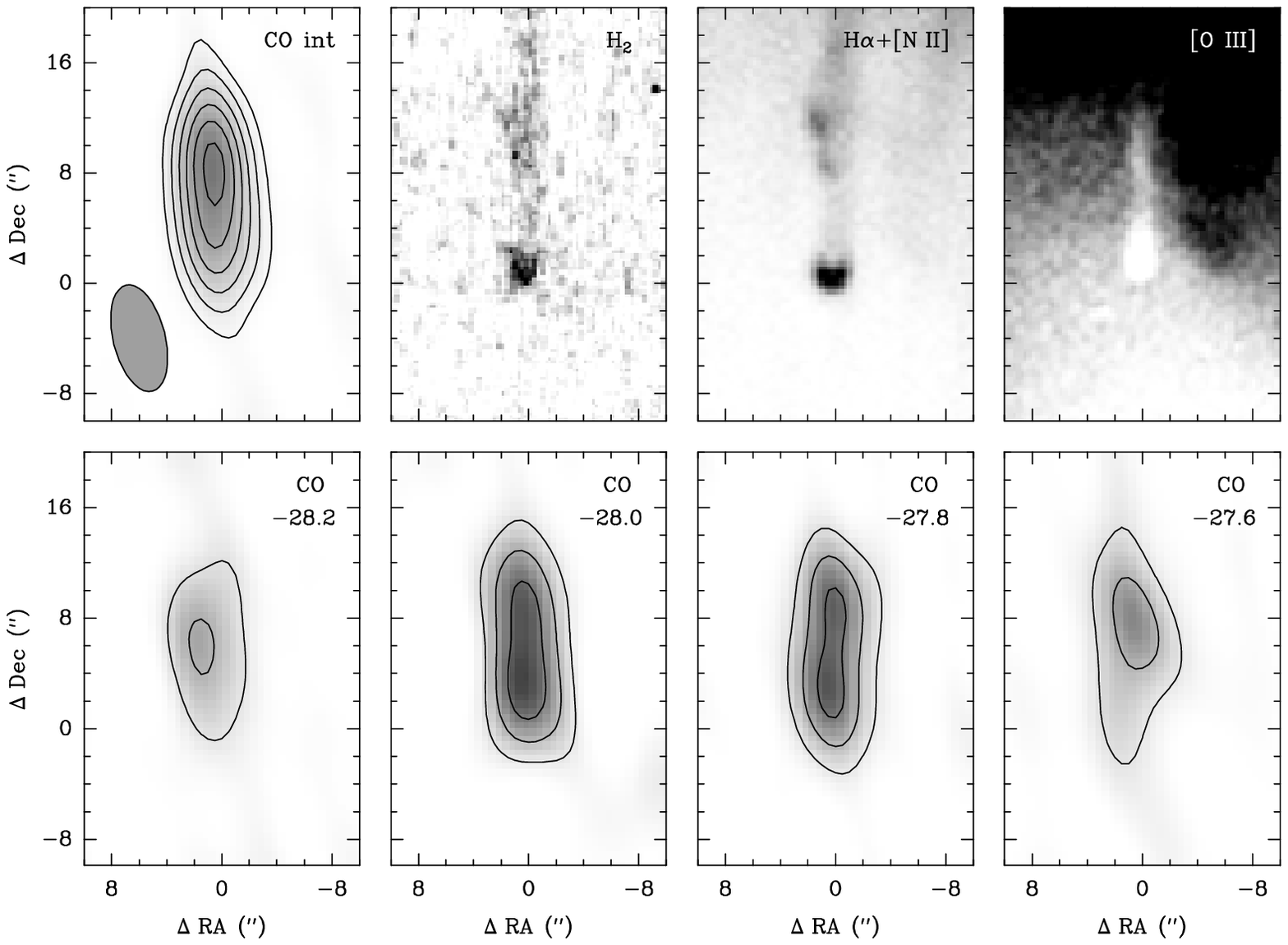}
\end{figure*}

\clearpage

\begin{figure}
\epsscale{0.5}
\plotone{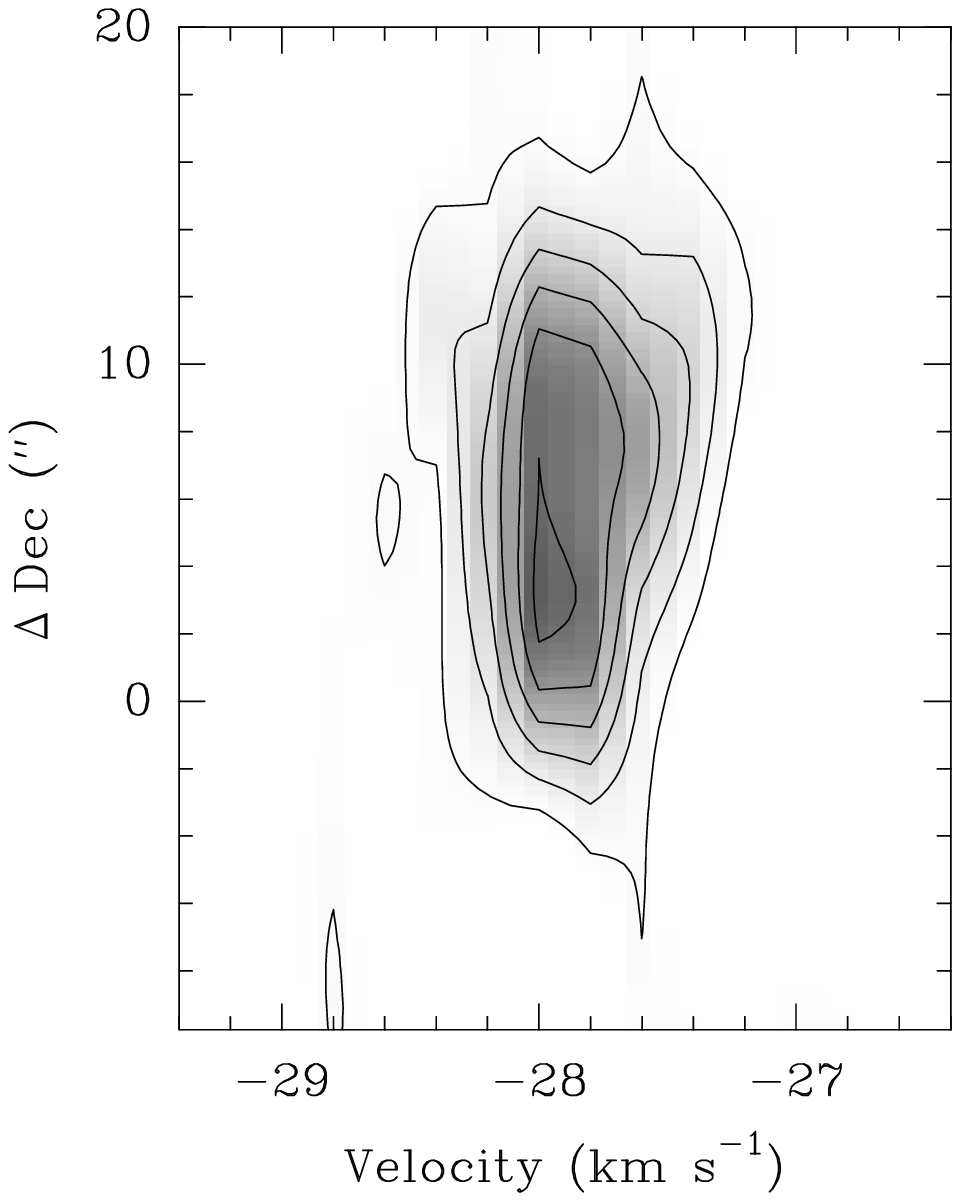}
\end{figure}

\clearpage

\begin{figure}
\epsscale{0.6}
\plotone{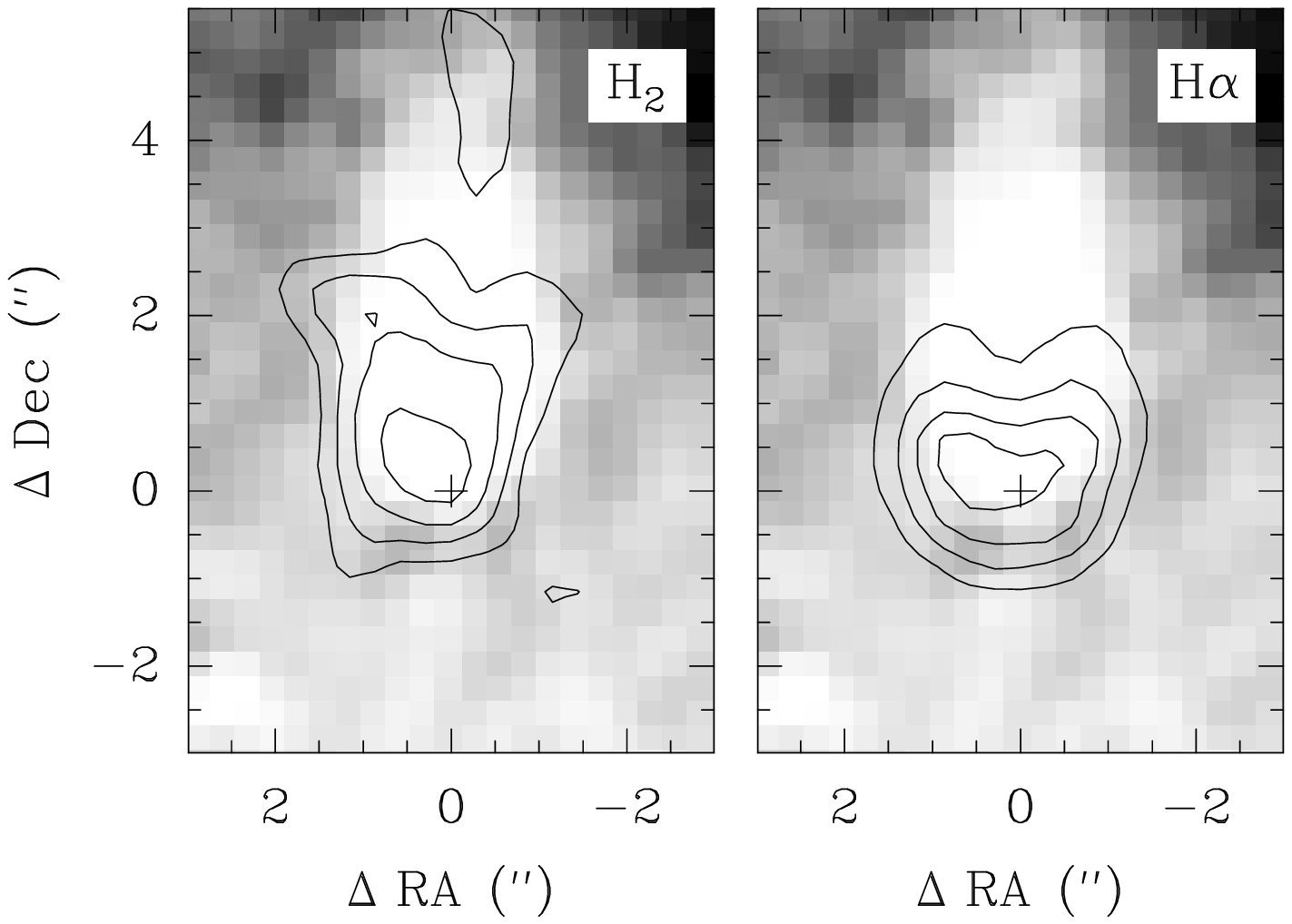}
\end{figure}

\end{document}